\documentclass[10pt,conference]{IEEEtran}  

\IEEEoverridecommandlockouts                              

\PassOptionsToPackage{dvipsnames,table}{xcolor}
\usepackage[dvipsnames,table]{xcolor}

\usepackage{tikz}
\usepackage{colortbl}
\usepackage{graphicx}
\usepackage{mathtools, cuted}
\usepackage{amsmath}
\usepackage{amssymb}
\usepackage{fancyhdr}
\usepackage{booktabs}
\usepackage{comment}
\usepackage{authblk}
\usepackage{graphics}

\usepackage{multicol}
\usepackage{multirow}
\usepackage[linesnumbered,ruled,vlined]{algorithm2e}
\SetKwRepeat{Do}{do}{while}%
\usepackage{amsmath,amssymb}
\usepackage[compatibility=false]{caption}
\usepackage[caption=false]{subfig}
\usepackage{paralist}
\usepackage{dblfloatfix}
\usepackage{subcaption}
\usepackage{setspace}
\usepackage{wrapfig}
\usepackage[hyphens]{url}
\usepackage{amsthm}
\usepackage{relsize}
\usepackage{amssymb}
\usepackage{pifont}
\usepackage{graphicx}
\usepackage{threeparttable}
\usepackage{booktabs}
\usepackage{dblfloatfix}

\newcommand{\polarai}{\texttt{POLARIS}}

\usepackage[dvipsnames]{xcolor}
\usepackage{tikz}
\newcommand{\tikzcircle}[2][red,fill=red]{\tikz[baseline=-0.5ex]\draw[#1,radius=#2] (0,0) circle ;}%
\makeatletter
\newcommand{\multiline}[1]{%
  \begin{tabularx}{\dimexpr\linewidth-\ALG@thistlm}[t]{@{}X@{}}
    #1
  \end{tabularx}
}
\makeatother

\title{POLARIS: Explainable Artificial Intelligence for Mitigating Power Side-Channel Leakage}

\author[1]{Tanzim Mahfuz}
\author[2]{Sudipta Paria}
\author[1]{Tasneem Suha}
\author[2]{Swarup Bhunia}
\author[1]{Prabuddha Chakraborty}
\affil[1]{Department of Electrical \& Computer Engineering, University of Maine, Orono, ME, USA}
\affil[2]{Department of Electrical \& Computer Engineering, University of Florida, Gainesville, FL, USA}

\begin{document}

\maketitle
\thispagestyle{fancy}
\pagestyle{empty}
\fancyhf{}
\fancyfoot[C]{\large \textbf{ \textcolor{blue}{© This paper has been accepted at the 62\textsuperscript{nd} Design Automation Conference (DAC), 2025}}}

\begin{abstract}
Microelectronic systems are widely used in many sensitive applications (e.g., manufacturing, energy, defense). These systems increasingly handle sensitive data (e.g., encryption key) and are vulnerable to diverse threats, such as, power side-channel attacks, which infer sensitive data through dynamic power profile. In this paper, we present a novel framework, \polarai~for mitigating power side channel leakage using an Explainable Artificial Intelligence (XAI) guided masking approach. \polarai~uses an unsupervised process to automatically build a tailored training dataset and utilize it to train a masking model. The \polarai~framework outperforms state-of-the-art mitigation solutions (e.g., VALIANT) in terms of leakage reduction, execution time, and overhead across large designs.

\end{abstract}



\section{Introduction}
Edge computing devices in Internet-of-Things (IoT) applications are increasingly used in diverse applications from domains, such as, Industry 4.0, surveillance, smart cities, defense, healthcare, and aerospace. These devices often record and process sensitive data and hence become a target for diverse attacks from malicious entities. A malicious entity can attempt to leak sensitive information from these devices through various means, such as timing side-channel attacks, power side-channel attacks, and fault-injection attacks. 

In this work, we focus on mitigating power side-channel attacks \cite{kocher,moradi_side_channel}, which involve measuring and analyzing the power consumption traces of a system to infer the values of sensitive data. This inference is possible because different computing instructions consume different amount of power while running on an unprotected/unmasked electronic hardware. Wide array of techniques have been developed to mitigate this concern, which include:
\begin{enumerate}
    \item \textbf{Quantifying Power Side Channel Leakage}: Techniques such as Test Vector Leakage Assessment (TVLA) were developed to estimate power side channel leakage at a gate level for a certain number of traces \cite{tvla}. 
    \item \textbf{Masking Gates}: Composite logic gates were developed to replace traditional gates to provide better power side channel leakage camouflaging \cite{trichina, dom}.   
    \item \textbf{Efficient Utilization of Masking Gates}: Brute force replacement of all gates using composite masking gates can lead to very high design overheads (area, power, delay). Hence, techniques such as VALIANT \cite{valiant} were developed to perform selective masking to obtain a good masking effect at a lower overhead. 
\end{enumerate}

Existing power side-channel countermeasures, such as, \cite{valiant,counter_3,karna}, which relies on fixed heuristics (static algorithm), may not perform well for diverse designs. Additionally, relying on TVLA (a simulation-based approach) analysis makes it highly time-consuming to operate on large designs. DL \cite{dl_la} and LLM-based \cite{netlist_whisper} leakage estimation techniques are proposed but are inadequate in leakage mitigation due to their high training times and lack of synthetic data support or explainability features. To mitigate these shortcomings, we present a novel Explainable AI based Design-for-Security framework \polarai~(\underline{PO}wer Side-Channel \underline{L}e\underline{A}kage \underline{R}eduction using Explainable A\underline{I} \underline{S}olution), that utilizes a search process to automatically learn how to most optimally insert masking gates to achieve high power side channel leakage reduction at low overhead. We implement \polarai~with training data generated using smaller open-source designs and extensively evaluated its effectiveness across various designs with varying complexities. \polarai~is able to outperform VALIANT \cite{valiant} in terms of performance (execution time), effectiveness (overall leakage reduction), and design overhead (area, power, delay). \polarai~does not rely on TVLA for leakage estimation and mitigation through masking and is approximately $6\mathsf{x}$ faster than VALIANT, making it scalable for larger designs. In summary, we make the following key contributions:

\begin{enumerate}
    \item Developed a novel explainable design-for-security (DFS) framework (\polarai) for mitigating power side-channel vulnerabilities in electronic systems. 
    \item Designed an unsupervised algorithm to automatically generate training data for \polarai~using a search approach and an AI algorithm for efficiently masking a given digital design.
    \item Implemented the \polarai~framework as a parameterized tool \& integrated it into the ASIC design flow. 
    \item Extensively evaluated \polarai~(different settings) and compared it with VALIANT \cite{valiant}, a state-of-the-art alternative solution, using more than 10 large digital designs.   
\end{enumerate}

The rest of the paper is organized as follows: Section II describes the brief background and related work, followed by the motivation behind this work in Section III. Section IV outlines the methodology, and Section V demonstrates the experimental results. We conclude the paper in Section VI.

\section{Background}\label{sec:relWorks} 

\subsection{Test Vector Leakage Assessment (TVLA)}
TVLA \cite{tvla} is a widely recognized method for power side-channel analysis based on Welch's t-test. The approach captures the DUT’s leakage behavior under varied inputs to detect any differences. Mathematically, let us denote two sets of data by $\mathcal{Q}_0$ and $\mathcal{Q}_1$, their cardinality by $n_0$ and $n_1$, respectively. Let also $\mu_0$ (resp. $\mu_1$) and $s_0^2$ (resp. $s_1^2$) denote sample mean and sample variance of the set $\mathcal{Q}_0$ (resp. $\mathcal{Q}_1$). The $t$-test statistic and the degree of freedom $v$ are computed as:
\vspace{-1em}
\begin{equation}
    t = \frac{\mu_0-\mu_1}{\sqrt{(\frac{s_0^2}{n_0})+(\frac{s_1^2}{n_1})}} \hspace{3em}
    v = \frac{{(\frac{s_0^2}{n_0}+\frac{s_1^2}{n_1})}^2}{\frac{{(\frac{s_0^2}{n_0})}^2}{n_0 - 1} + \frac{{(\frac{s_1^2}{n_1})}^2}{n_1 - 1}} 
\end{equation}


The t-statistic is typically assessed with a threshold of $\pm$ 4.5 for distinguishability. This threshold ensures that for $\left|t\right| > $ 4.5 and $v > $ 1000, the p-value drops below 0.00001, indicating over 99.999\% confidence against the null hypothesis \cite{moradi}. A crypto implementation is considered protected with 99.999\% confidence if $\left|t\right| \leq 4.5$; otherwise, it may indicate leakage \cite{valiant}. The test can be performed in two ways: (i) Fixed-vs-Fixed (using known intermediate values) or (ii) Fixed-vs-Random (using fixed and random input patterns) \cite{masking_flaw}.

The naive algorithm to compute variance ($s_0^2,s_1^2$) followed by the respective standard deviation ($\rho_0,\rho_1$) is inefficient as it involves processing full trace points twice due to the dependency on mean ($\mu_0,\mu_1$) values for the computation. The evaluation for $\rho_0$ is given as:
\vspace{-0.5em}
\begin{equation}
    \rho_0 = \frac{\sum_{i}{(x_i - \mu_0)}^2}{n_0}, \text{ where } x_i \in \mathcal{Q}_0
\end{equation}


TVLA trace collection is slow due to repeated mean and variance calculations. To accelerate it, \cite{moradi} proposed an efficient one-pass method for raw and central moments computation during trace acquisition. The raw moment (sample mean) and central moment (variance) of order $d=1$ for an expanded set $\mathcal{Q}' = \mathcal{Q} \cup y$ with $n$ elements are calculated as:
\vspace{-0.5em}
\begin{equation}
    M_1^{\mathcal{Q}'} =  M_1^{\mathcal{Q}} + \frac{\Delta}{n}, \text{ where   } \Delta = y -  M_1^{\mathcal{Q}}
\end{equation}
\vspace{-1em}
\begin{equation}
    \mu = M_1 \hspace{0.5em} \text{and} \hspace{0.5em} s^2 = CM_2,\text{ where  } CM_2 = M_2 - M_1^2
\end{equation}

This method can be extended to compute the central moments at any $d>$ 1 using the equations as given in \cite{moradi}.

\subsection{Masking}

\begin{figure}[!ht]
\centering
\vspace{-2em}
\includegraphics[width=\columnwidth]{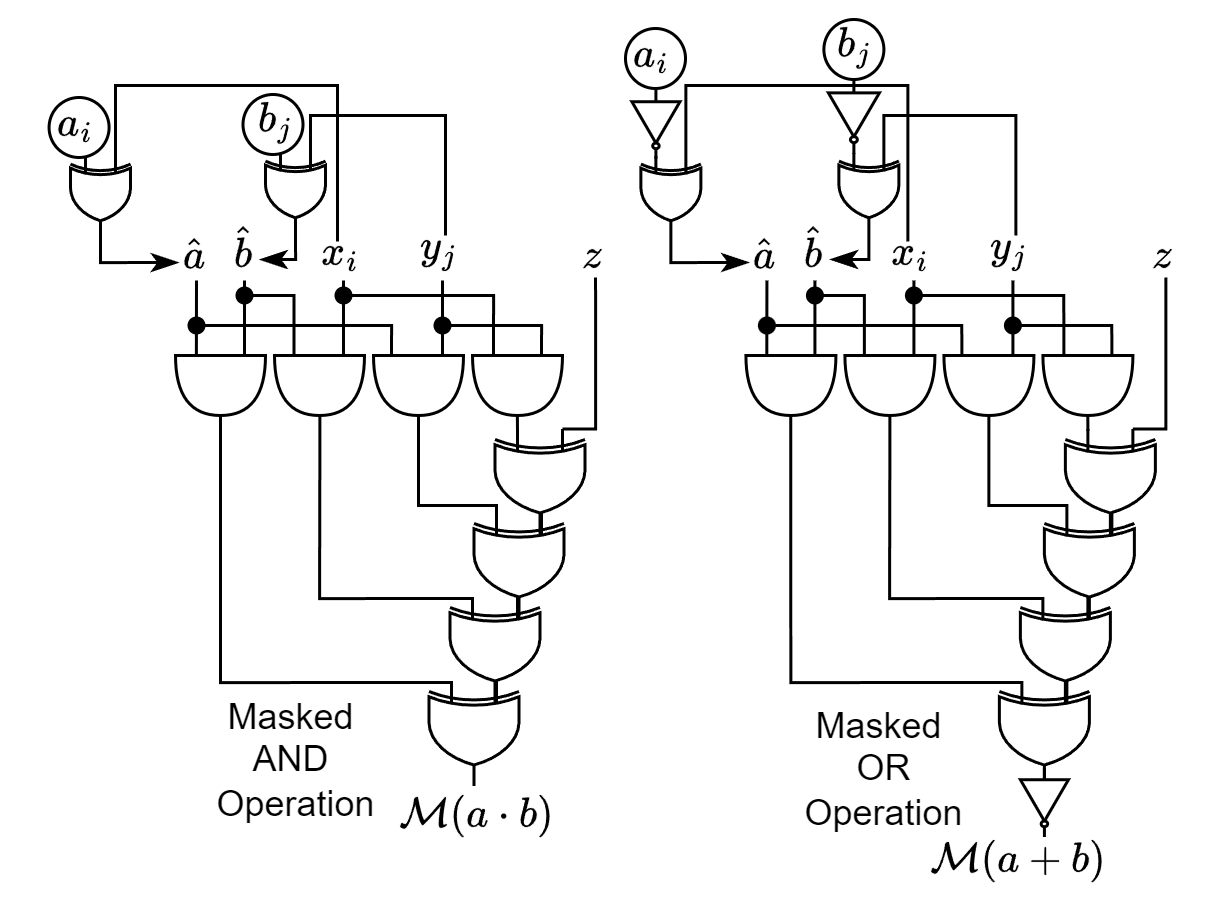}
\caption{\small{\textit{AND} and \textit{OR} operation on masked data. \label{fig:trichina}}}

\vspace{-1em}
\end{figure}

Masking is a well-known countermeasure for mitigating side-channel attacks \cite{masking_proof}. Masking randomizes sensitive data in cryptographic operations using a secret sharing scheme at the logic level. To achieve $d$-th order security, each variable $x$ is split randomly into $(d+1)$ shares, such that
$x = x^{(1)} \oplus x^{(2)} \oplus \cdots \oplus x^{(d+1)}$, where $x^{(i)}$ denotes $i^{th}$~share of $x$ and $\oplus$ denotes XOR operation. This provides significant protection since that adversary would need to recover all $d+1$ shares to obtain the actual value of $x$, which requires statistical analysis up to $d^{th}$ order. Authors in \cite{trichina,dom} proposed efficient masking schemes for block cipher implementations against DPA attacks. As example, the \textit{Masked} \textit{AND} and \textit{OR} operation proposed by \cite{trichina} is demonstrated in Fig. \ref{fig:trichina} and described below:

If $x_i$ and $y_j$ denote the bits that mask the `real' bits $a_i$ and $b_j$ then we denote masked bits as $\hat{a} = (a_i \oplus x_i)$ and $\hat{b} = (b_j \oplus y_j)$ correspondingly. We introduce a random bit, z, as a new mask for computing a \textit{masked} \textit{AND} operation as: 
\vspace{-0.5em}
\begin{equation}
    \mathcal{M}(a \cdot b) = ((\hat{a} \cdot \hat{b}) \oplus ((x_i \cdot \hat{b}) \oplus ((x_i \cdot y_j ) \oplus z))) \oplus (y_j \cdot \hat{a})
\end{equation}

\begin{figure*}[!ht]
\centering

\includegraphics[width=0.8\linewidth]{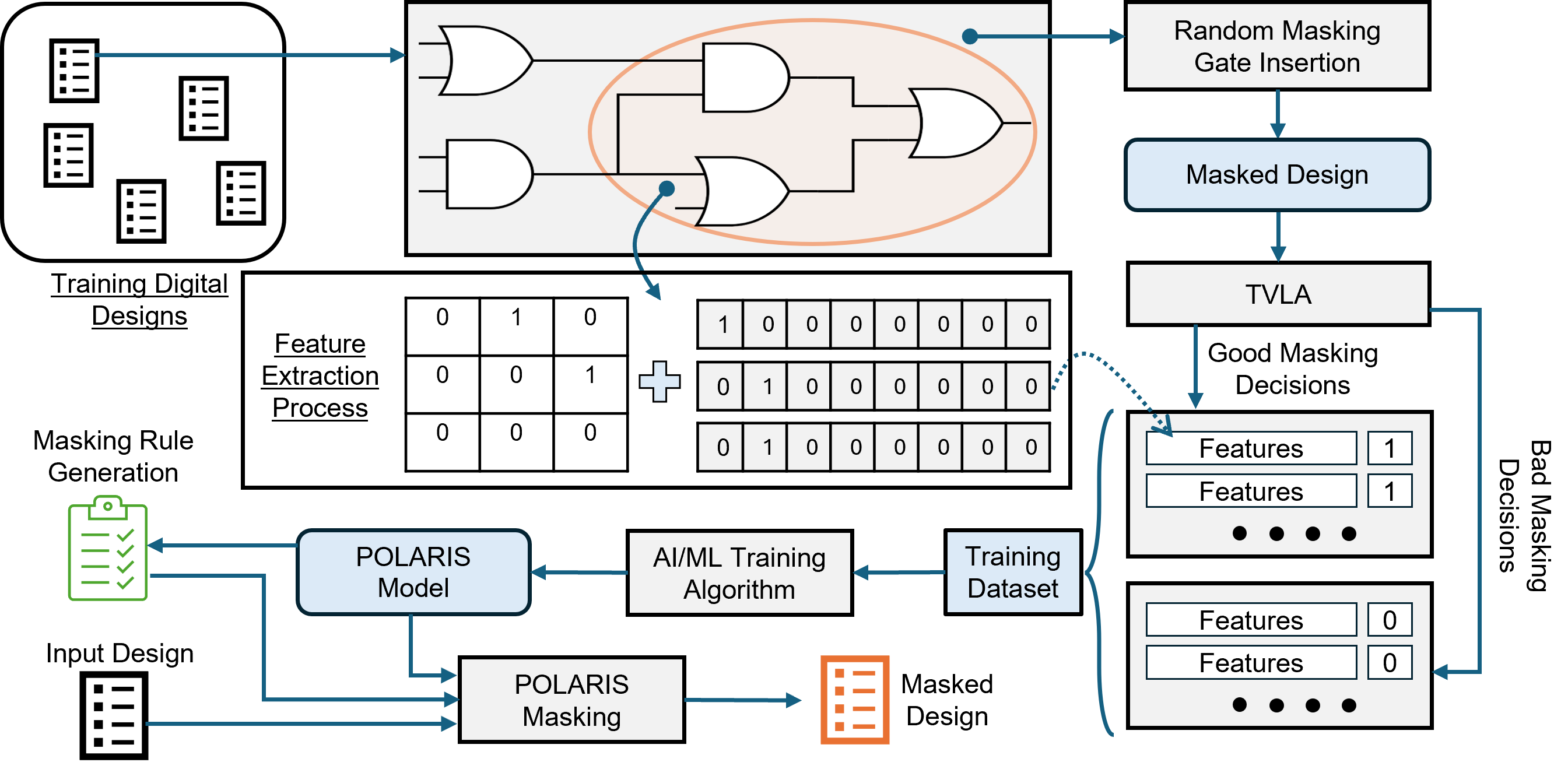}
\caption{\small{Overview of the \polarai~framework. \label{fig:overview}}}

\vspace{-0.2in}
\end{figure*}

\begin{table*}[!hb]
\centering
\caption{Comparing \polarai~with existing solutions for power side-channel leakage assessment and mitigation.}
\label{table:background}


\resizebox{\textwidth}{!}{%
\begin{tabular}{|c|c|cc|c|c|c|c|}
\hline
\rowcolor[HTML]{ECF4FF} 
\textbf{Approach}                                                                      & \textbf{Method} & \multicolumn{2}{c|}{\cellcolor[HTML]{ECF4FF}\textbf{Model} \textbf{Training}}                                                                                                                 & \textbf{\begin{tabular}[c]{@{}c@{}}Feature \\ Set\end{tabular}}                                          & \textbf{\begin{tabular}[c]{@{}c@{}}Mitigation \\ Support\end{tabular}} & \textbf{Performance} & \textbf{Platform}                                     \\ \hline
\cellcolor[HTML]{EDF9FA}\textbf{CASCADE} \cite{counter_3}                                              & TVLA            & \multicolumn{2}{c|}{N/A}                                                                                                                                      & N/A                                                            & No                                                                     & Slow                 & ASIC                                                  \\ \hline
\cellcolor[HTML]{EDF9FA}\textbf{Karna} \cite{karna}                                                & TVLA            & \multicolumn{2}{c|}{N/A}                                                                                                                                      & N/A                                                            & Limited                                                                & Slow                 & ASIC                                                  \\ \hline
\cellcolor[HTML]{EDF9FA}\textbf{VALIANT}  \cite{valiant}                                             & TVLA            & \multicolumn{2}{c|}{N/A}                                                                                                                                      & N/A                                                            & Yes                                                                    & Slow                 & ASIC                                                  \\ \hline
\cellcolor[HTML]{EDF9FA}\begin{tabular}[c]{@{}c@{}}\textbf{DL-LA} \cite{dl_la}\end{tabular}                                                & DL              & \multicolumn{1}{c|}{\begin{tabular}[c]{@{}c@{}}Training Time: High\\ Adversarial Attacks: Possible\end{tabular}} & \begin{tabular}[c]{@{}c@{}}Explainability: No \\ Synthetic Data Support: No\end{tabular} & \begin{tabular}[c]{@{}c@{}}Trace \\ based\end{tabular}         & No                                                                     & Slow                 & \begin{tabular}[c]{@{}c@{}}ASIC/\\ FPGA\end{tabular}  \\ \hline
\cellcolor[HTML]{EDF9FA}\begin{tabular}[c]{@{}c@{}}\textbf{Netlist} \\ \textbf{Whisperer} \cite{netlist_whisper}\end{tabular}                                                & LLM              & \multicolumn{1}{c|}{\begin{tabular}[c]{@{}c@{}}Training Time: High\\ Adversarial Attacks: Possible\end{tabular}} & \begin{tabular}[c]{@{}c@{}}Explainability: No \\ Synthetic Data Support: No\end{tabular} & \begin{tabular}[c]{@{}c@{}}ANF \\ equations\end{tabular}         & Yes                                                                     & Slow                 & \begin{tabular}[c]{@{}c@{}}ASIC\end{tabular}  \\ \hline
\cellcolor[HTML]{EDF9FA}\textbf{\begin{tabular}[c]{@{}c@{}} \polarai \\(This work) \end{tabular}} & XAI             & \multicolumn{1}{c|}{\begin{tabular}[c]{@{}c@{}}Training Time: Low\\ Adversarial Attacks: No\end{tabular}}        & \begin{tabular}[c]{@{}c@{}}Explainability: Yes \\ Synthetic Data Support: Yes\end{tabular} & \begin{tabular}[c]{@{}c@{}}Structural \\ Analysis\end{tabular} & Yes                                                                    & Fast                 & \begin{tabular}[c]{@{}c@{}}ASIC/\\ FPGA*\end{tabular} \\ \hline \hline
\end{tabular}
}

\footnotesize{\textbf{*}\polarai~can be extended to FPGA design flow by re-training the model with lookup table (LUT) based FPGA netlists.} \raggedright
\end{table*}

\subsection{Related Work}

Existing power side-channel leakage evaluation/mitigation approaches like CASCADE \cite{counter_3}, Karna \cite{karna}, VALIANT \cite{valiant} incorporate TVLA analysis and hence suffers from high runtime and scalability issues. AI-based methods like DL-LA \cite{dl_la} and Netlist Whisperer \cite{netlist_whisper} are often limited by the lack of training data and are vulnerable to adversarial attacks due to reliance on external training data \cite{poison?,poison,backdoorai,backdoor}.

\polarai~stands out by using a synthetic data generation approach for creating its own training data, ensuring improved performance and resilience against adversarial attacks. \polarai~also incorporates XAI-based rules generation for efficient power side channel mitigation through masking. Table \ref{table:background} provides a comparative analysis between the existing solutions and the proposed framework.


\vspace{-0.5em}
\section{Motivation}
\label{sec:motivation}
\subsection{Can Explainable AI Generate Custom DFS Rules/Models for Power Side Channel Mitigation?}
Power side channel mitigation techniques such as Karna and VALIANT rely on static algorithms, in other words, a fixed set of hand-crafted rules. This is a limiting factor and prevents optimal operation on a diverse set of designs. We hypothesize that an Explainable AI (XAI) framework might be able to generate custom rules/models (to mitigate power side channel leakage) based on a given type of design, making the process more autonomous and adaptive.   

\subsection{Can We Bypass TVLA using AI?} Most of the prevalent power side channel mitigation frameworks (e.g., Karna, VALIANT) rely on TVLA for estimating power leakage. TVLA has major scalability concerns particularly for large industry-grade designs that in turn hampers the scalability of frameworks such as Karna and VALIANT. We hypothesize, AI techniques might be able to predict the leakage (directly or in an integrated way) associated with individual gates faster and with a linear time complexity (with respect to design size).  

\subsection{Can We Generate the Training Data?}
Training data is a major limiting factor for AI frameworks in the hardware security domain. Techniques such as DL-LA \cite{dl_la} and Netlist Whisperer \cite{netlist_whisper} require a very high amount of data making it almost infeasible to effectively scale across different design types. If we can develop a framework that can leverage synthetic training data automatically generated via an unsupervised technique, then it would be highly scalable across diverse design types. 
 
\section{Methodology}\label{sec:method}

We build on the motivations presented in Sec.~\ref{sec:motivation} to design a framework that can: (1) Leverage XAI for creating custom power side channel mitigation rules/models; (2) Bypass TVLA using an AI approach; (3) Utilize automatically generated synthetic data for model training.   
The overall framework, shown in Fig.~\ref{fig:overview} operates in three major stages: (i) Knowledge extraction, data generation, and building the ML model; (ii) Interpreting the model using the explainable SHAP \cite{lundberg2017unified} framework; and (iii) Masking against side-channel vulnerabilities. 

\subsection{Knowledge Extraction and ML Model Development}

Obtaining a large dataset for training AI models is a major concern particularly in the domain of hardware security and EDA. To address this concern, \polarai~utilizes a novel unsupervised approach to generate a large database for AI model training. As shown in Fig.~\ref{fig:overview}, the framework \polarai~converts any digital design represented as gate-level netlist ($\mathcal{D}$) into a graph ($G_r$) such that $G_r = (V,E)$ where, $V$: gates and $E$: interconnections and randomly inserts masking gates based on mask size ($\mathcal{M}_{size}$). Then, \polarai~calculates leakage values ($\mathbb{L}_G$) using TVLA analysis and compares it to the original leakage values. If the difference exceeds a threshold ($\theta_r$), the masking is labeled `good' ($1$); otherwise, `bad' ($0$) and appended to $\{X\_data, Y\_data\}$. These labels are linked to structural features used for knowledge extraction. The framework employs local structural features for training and evaluation, similar to \cite{chakraborty2018sail},\cite{chakraborty2024learning}. The structural features of a gate include information such as their local placement and interconnections. In a sub-design graph, gate connectivity is encoded with an adjacency matrix and one-hot encoding. Fig.~\ref{fig:overview} illustrates the structural feature extraction process, showing how gates are vectorized and how the dataset is created via feature extraction and labeling. Breadth-first search (BFS) is employed to explore neighboring gates (Locality $\mathcal{L}$) and assign labels based on leakage comparison. The random insertion process runs iteratively ($\leq itr$ times) until the termination condition is met. Finally, we obtain a trained model ($\mathbb{M}$) trained using $\{X\_data, Y\_data\}$  that will be used to generate human-understandable rules through model interpretability in the next stage. Algorithm~\ref{algo:knowledge_extraction} describes the sequence of steps involved in extracting knowledge through structural features and developing the ML model.
\setlength{\textfloatsep}{0pt}
\begin{algorithm}[!t]
\DontPrintSemicolon 
\KwIn{$\mathcal{D}, \mathcal{M}_{size}, \mathcal{L}, itr, \theta_{r}$}
\KwOut{$\mathbb{M}$}

$G_r \gets$ \textbf{\textit{graphify}}($\mathcal{D}$) \;
$\mathbb{L}_G \gets$ \textbf{\textit{leak\_estimate}}($\mathcal{D}$) \;
$X\_data, Y\_data \gets \emptyset $; $run \gets 0$ \;
$\mathcal{R}_{gates} \gets G_r.gates$ \;

\While{$\mathcal{M}_{size} \le len(\mathcal{R}_{gates})$ and $run \le itr$}
{
   $\mathcal{S}_{gates} \gets$ \textbf{\textit{random}}($\mathcal{M}_{size}, \mathcal{R}$)\;
   $\mathcal{D}_{mod} \gets$ \textbf{\textit{modify}}($\mathcal{S}_{gates},\mathcal{D}$)\;
    $\mathcal{R}_{gates} \gets \mathcal{R}_{gates} - \mathcal{S}_{gates}$\;
    $\mathbb{L}_G^{mod} \gets$ \textbf{\textit{leak\_estimate}}($\mathcal{D}_{mod}$) \;
    
    \For{$i$ \textbf{in} $\mathcal{S}_{gates}$}
    {
        $S_f \gets$ \textbf{\textit{structural\_features}}($G_r,\mathcal{L},i$)\;
        $r_{Ratio} \gets$ \textbf{\textit{compare}}($\mathbb{L}_G[i],\mathbb{L}_G^{mod}[i]$)\;
        \If{$r_{Ratio} \ge \theta_{r}$} 
        {
            $label \gets 1$\;
        }
        \Else
        {
             $label \gets 0$\;
        }

        $X\_data.append(S_f)$ \;
        $Y\_data.append(label)$ \;
    }
    $run \gets run + 1$ \;
    
}
$\mathbb{M} \gets$ \textbf{\textit{train\_data}}($X\_data$,$Y\_data$)\;
\Return{$\mathbb{M}$}\;
\caption{Cognition Generation}
\label{algo:knowledge_extraction}
\end{algorithm}



\subsection{Explainable AI (XAI): Model Interpretability by SHAP}
XAI enhances transparency and accountability in AI by providing methods for designing inherently interpretable models and for analyzing decisions after training. We incorporate the SHAP (SHapley Additive exPlanations) \cite{lundberg2017unified} algorithm into our framework to analyze the trained models and extract the underlying model rules driving the mitigation process. 
SHAP leverages game theory to quantify the contribution of each feature in model's prediction. For an individual prediction, the Shapley value of a feature \( f \), is denoted as \( \phi_f \), and is computed using Equation~\ref{eqn_shapley}. Given \( h \) total features, there are \( h! \) possible permutations (coalitions) of these features. Let \( g \) represent a specific coalition that excludes feature \( f \). The weighting factor \( \frac{|g|! (h - |g| - 1)!}{h!} \) accounts for the number of ways to form such coalitions, and the term \( \text{val}(g \cup \{ f \}) - \text{val}(g) \) signifies the marginal contribution of feature \( f \) to coalition \( g \). $N$ denotes set of all features $\{1,2,...h\}$. 

While there are other XAI frameworks, such as LIME \cite{ribeiro2016should} and Captum \cite{kokhlikyan2020captum}, we focus on SHAP due to its versatility. SHAP offers both model-agnostic methods (e.g., Kernel SHAP) and model-specific methods (e.g., Tree SHAP), making it suitable for a broad range of models. The automated rules, unlike handcrafted ones, can be used independently to make masking decisions or alongside the model to achieve better predictions and decisions, as shown in Fig.~\ref{fig:overview}.
\begin{equation}
\label{eqn_shapley}
\phi_f = \sum_{g \subseteq N \setminus \{ f \}} \frac{|g|! \left( h-|g|-1 \right)!}{h!} \left[ \text{val}\left( g \cup \{ f \} \right) - \text{val}(g) \right]
\end{equation}


\vspace{-1em}
\subsection{\polarai~Masking in Side-Channel Defense}



The trained model can be used directly to mask a design for side-channel defense and can also leverage human-readable rules ($\mathcal{R}_L$) rules generated through the XAI framework to improve decision-making. For each gate in the graph ($G_r$), generated from ($\mathcal{D}$), structural features ($S_f$) are extracted, and predictions ($\mathcal{P}_{pred}$) are made using the model or rules followed by appending $\mathcal{P}_{pred}$ values to choices ($\mathcal{C}$). The set $\mathcal{C}$ is sorted in descending order to identify the top selections ($\mathcal{C}_{top}$). Finally, masking is applied to the design based on $\mathcal{C}_{top}$, and leakage of the masked design ($\mathcal{D}_\mathcal{M}$) is measured. Algorithm~\ref{algo:xdfs_masking} outlines all the steps involved. 




\setlength{\textfloatsep}{0pt}
\begin{algorithm}[!t]
\DontPrintSemicolon 
\KwIn{$\mathcal{D}, \mathbb{M}, \mathcal{R}_{L}, \mathcal{M}_{size}, \mathcal{L}$}
\KwOut{$\mathcal{D_M}$}

$G_r \gets$ \textbf{\textit{graphify}}($\mathcal{D}$) \;
$\mathcal{G} \gets G_r.gates$ \;
$\mathcal{C} \gets \emptyset$ \;

\For{$i$ \textbf{in} $\mathcal{G}$}
{
    $S_f \gets$ \textbf{\textit{structural\_features}}($G_r,\mathcal{L},i$)\;
    $\mathcal{P}_{pred} \gets$ \textbf{\textit{inferencing}}($\mathbb{M}, R_L, S_f$)\;
     $\mathcal{C}.append(\mathcal{P}_{pred},i)$\;
}
$\mathcal{C}_{top} \gets$ \textbf{\textit{sort\_descending}}($\mathcal{C}$)\;
$\mathcal{D_M} \gets$ \textbf{\textit{modify}}($\mathcal{D}, \mathcal{C}_{top}, \mathcal{M}_{size}$)\;
$\mathbb{L}_{info} \gets$ \textbf{\textit{leak\_estimate}}($\mathcal{D_M}$) \;
\Return{$\mathcal{D_M}$}\;
\caption{\polarai~Masking}
\label{algo:xdfs_masking}
\end{algorithm}

\begin{table*}[!htbp]
\centering
\captionsetup{justification=centering}
\caption{Quantitative comparison between VALIANT \cite{valiant} \& \polarai~in terms of leakage reduction \& runtime efficiency.}
\label{table:core_result}



\begin{threeparttable}
\resizebox{\textwidth}{!}{%
\begin{tabular}{|c|ccccc|cccc|cc|}
\hline
\rowcolor[HTML]{ECF4FF} 
\cellcolor[HTML]{ECF4FF}                                      & \multicolumn{5}{c|}{\cellcolor[HTML]{ECF4FF}\textbf{Leakage Value (Per Gate)}}                                                                                                                                                                                                                                            & \multicolumn{4}{c|}{\cellcolor[HTML]{ECF4FF}\textbf{Total Leakage Reduction (\%)}}                                                                                                                                                       & \multicolumn{2}{c|}{\cellcolor[HTML]{ECF4FF}\textbf{Time (s)}}                                                                                                                             \\ \cline{2-12} 
\rowcolor[HTML]{EFEFEF} 
\cellcolor[HTML]{ECF4FF}                                      & \multicolumn{1}{c|}{\cellcolor[HTML]{EFEFEF}}                                  & \multicolumn{1}{c|}{\cellcolor[HTML]{EFEFEF}}                                   & \multicolumn{3}{c|}{\cellcolor[HTML]{EFEFEF}\textbf{\polarai}}                                                                                       & \multicolumn{1}{c|}{\cellcolor[HTML]{EFEFEF}}                                   & \multicolumn{3}{c|}{\cellcolor[HTML]{EFEFEF}\textbf{\polarai}}                                                                                       & \multicolumn{1}{c|}{\cellcolor[HTML]{EFEFEF}}                                   & \cellcolor[HTML]{EFEFEF}                                                                                 \\ \cline{4-6} \cline{8-10}
\rowcolor[HTML]{EFEFEF} 
\multirow{-3}{*}{\cellcolor[HTML]{ECF4FF}\textbf{Benchmarks}} & \multicolumn{1}{c|}{\multirow{-2}{*}{\cellcolor[HTML]{EFEFEF}\textbf{Before}}} & \multicolumn{1}{c|}{\multirow{-2}{*}{\cellcolor[HTML]{EFEFEF}\textbf{VALIANT}}} & \multicolumn{1}{c|}{\cellcolor[HTML]{EFEFEF}\textbf{50\%Mask}} & \multicolumn{1}{c|}{\cellcolor[HTML]{EFEFEF}\textbf{75\%Mask}} & \textbf{Full Mask} & \multicolumn{1}{c|}{\multirow{-2}{*}{\cellcolor[HTML]{EFEFEF}\textbf{VALIANT}}} & \multicolumn{1}{c|}{\cellcolor[HTML]{EFEFEF}\textbf{50\%Mask}} & \multicolumn{1}{c|}{\cellcolor[HTML]{EFEFEF}\textbf{75\%Mask}} & \textbf{Full Mask} & \multicolumn{1}{c|}{\multirow{-2}{*}{\cellcolor[HTML]{EFEFEF}\textbf{VALIANT}}} & \multirow{-2}{*}{\cellcolor[HTML]{EFEFEF}\textbf{\begin{tabular}[c]{@{}c@{}}\polarai \end{tabular}}} \\ \hline
\rowcolor[HTML]{FFFFFF} 
\cellcolor[HTML]{EDF9FA}\textbf{des3}                         & \multicolumn{1}{c|}{\cellcolor[HTML]{FFFFFF}2.66}                              & \multicolumn{1}{c|}{\cellcolor[HTML]{FFFFFF}1.20}                               & \multicolumn{1}{c|}{\cellcolor[HTML]{FFFFFF}1.23}               & \multicolumn{1}{c|}{\cellcolor[HTML]{FFFFFF}1.09}               & 1.02               & \multicolumn{1}{c|}{\cellcolor[HTML]{FFFFFF}54.89}                              & \multicolumn{1}{c|}{\cellcolor[HTML]{FFFFFF}53.76}              & \multicolumn{1}{c|}{\cellcolor[HTML]{FFFFFF}59.02}              & 61.65              & \multicolumn{1}{c|}{\cellcolor[HTML]{FFFFFF}38.68}                              & 16.40                                                                                                    \\ \hline
\rowcolor[HTML]{FFFFFF} 
\cellcolor[HTML]{EDF9FA}\textbf{arbiter}                      & \multicolumn{1}{c|}{\cellcolor[HTML]{FFFFFF}1.42}                              & \multicolumn{1}{c|}{\cellcolor[HTML]{FFFFFF}1.10}                               & \multicolumn{1}{c|}{\cellcolor[HTML]{FFFFFF}1.15}               & \multicolumn{1}{c|}{\cellcolor[HTML]{FFFFFF}0.98}               & 0.88               & \multicolumn{1}{c|}{\cellcolor[HTML]{FFFFFF}22.54}                              & \multicolumn{1}{c|}{\cellcolor[HTML]{FFFFFF}19.01}              & \multicolumn{1}{c|}{\cellcolor[HTML]{FFFFFF}30.99}              & 38.03              & \multicolumn{1}{c|}{\cellcolor[HTML]{FFFFFF}97.62}                              & 15.11                                                                                                    \\ \hline
\rowcolor[HTML]{FFFFFF} 
\cellcolor[HTML]{EDF9FA}\textbf{sin}                          & \multicolumn{1}{c|}{\cellcolor[HTML]{FFFFFF}2.16}                              & \multicolumn{1}{c|}{\cellcolor[HTML]{FFFFFF}1.23}                               & \multicolumn{1}{c|}{\cellcolor[HTML]{FFFFFF}1.19}               & \multicolumn{1}{c|}{\cellcolor[HTML]{FFFFFF}1.04}               & 0.91               & \multicolumn{1}{c|}{\cellcolor[HTML]{FFFFFF}43.06}                              & \multicolumn{1}{c|}{\cellcolor[HTML]{FFFFFF}44.91}              & \multicolumn{1}{c|}{\cellcolor[HTML]{FFFFFF}51.85}              & 57.87              & \multicolumn{1}{c|}{\cellcolor[HTML]{FFFFFF}42.42}                              & 15.21                                                                                                    \\ \hline
\rowcolor[HTML]{FFFFFF} 
\cellcolor[HTML]{EDF9FA}\textbf{md5}                          & \multicolumn{1}{c|}{\cellcolor[HTML]{FFFFFF}2.93}                              & \multicolumn{1}{c|}{\cellcolor[HTML]{FFFFFF}1.39}                               & \multicolumn{1}{c|}{\cellcolor[HTML]{FFFFFF}1.41}               & \multicolumn{1}{c|}{\cellcolor[HTML]{FFFFFF}1.28}               & 1.21               & \multicolumn{1}{c|}{\cellcolor[HTML]{FFFFFF}52.56}                              & \multicolumn{1}{c|}{\cellcolor[HTML]{FFFFFF}51.88}              & \multicolumn{1}{c|}{\cellcolor[HTML]{FFFFFF}56.31}              & 58.70              & \multicolumn{1}{c|}{\cellcolor[HTML]{FFFFFF}202.11}                             & 59.11                                                                                                    \\ \hline
\rowcolor[HTML]{FFFFFF} 
\cellcolor[HTML]{EDF9FA}\textbf{voter}                        & \multicolumn{1}{c|}{\cellcolor[HTML]{FFFFFF}2.96}                              & \multicolumn{1}{c|}{\cellcolor[HTML]{FFFFFF}1.30}                               & \multicolumn{1}{c|}{\cellcolor[HTML]{FFFFFF}1.12}               & \multicolumn{1}{c|}{\cellcolor[HTML]{FFFFFF}1.10}               & 1.01               & \multicolumn{1}{c|}{\cellcolor[HTML]{FFFFFF}56.08}                              & \multicolumn{1}{c|}{\cellcolor[HTML]{FFFFFF}62.16}              & \multicolumn{1}{c|}{\cellcolor[HTML]{FFFFFF}62.84}              & 65.88              & \multicolumn{1}{c|}{\cellcolor[HTML]{FFFFFF}84.59}                              & 28.36                                                                                                    \\ \hline
\rowcolor[HTML]{FFFFFF} 
\cellcolor[HTML]{EDF9FA}\textbf{square}                       & \multicolumn{1}{c|}{\cellcolor[HTML]{FFFFFF}2.94}                              & \multicolumn{1}{c|}{\cellcolor[HTML]{FFFFFF}1.23}                               & \multicolumn{1}{c|}{\cellcolor[HTML]{FFFFFF}1.29}               & \multicolumn{1}{c|}{\cellcolor[HTML]{FFFFFF}1.25}               & 1.21               & \multicolumn{1}{c|}{\cellcolor[HTML]{FFFFFF}58.16}                              & \multicolumn{1}{c|}{\cellcolor[HTML]{FFFFFF}56.12}              & \multicolumn{1}{c|}{\cellcolor[HTML]{FFFFFF}57.48}              & 58.84              & \multicolumn{1}{c|}{\cellcolor[HTML]{FFFFFF}309.00}                             & 56.62                                                                                                    \\ \hline
\rowcolor[HTML]{FFFFFF} 
\cellcolor[HTML]{EDF9FA}\textbf{sqrt}                         & \multicolumn{1}{c|}{\cellcolor[HTML]{FFFFFF}3.04}                              & \multicolumn{1}{c|}{\cellcolor[HTML]{FFFFFF}1.44}                               & \multicolumn{1}{c|}{\cellcolor[HTML]{FFFFFF}1.48}               & \multicolumn{1}{c|}{\cellcolor[HTML]{FFFFFF}1.31}               & 1.20               & \multicolumn{1}{c|}{\cellcolor[HTML]{FFFFFF}52.63}                              & \multicolumn{1}{c|}{\cellcolor[HTML]{FFFFFF}51.32}              & \multicolumn{1}{c|}{\cellcolor[HTML]{FFFFFF}56.91}              & 60.53              & \multicolumn{1}{c|}{\cellcolor[HTML]{FFFFFF}281.05}                             & 75.53                                                                                                    \\ \hline
\rowcolor[HTML]{FFFFFF} 
\cellcolor[HTML]{EDF9FA}\textbf{div}                          & \multicolumn{1}{c|}{\cellcolor[HTML]{FFFFFF}2.28}                              & \multicolumn{1}{c|}{\cellcolor[HTML]{FFFFFF}1.44}                               & \multicolumn{1}{c|}{\cellcolor[HTML]{FFFFFF}1.37}               & \multicolumn{1}{c|}{\cellcolor[HTML]{FFFFFF}1.28}               & 1.18               & \multicolumn{1}{c|}{\cellcolor[HTML]{FFFFFF}36.84}                              & \multicolumn{1}{c|}{\cellcolor[HTML]{FFFFFF}39.91}              & \multicolumn{1}{c|}{\cellcolor[HTML]{FFFFFF}43.86}              & 48.25              & \multicolumn{1}{c|}{\cellcolor[HTML]{FFFFFF}389.26}                             & 92.51                                                                                                    \\ \hline
\rowcolor[HTML]{FFFFFF} 
\cellcolor[HTML]{EDF9FA}\textbf{memctrl}                      & \multicolumn{1}{c|}{\cellcolor[HTML]{FFFFFF}1.62}                              & \multicolumn{1}{c|}{\cellcolor[HTML]{FFFFFF}1.21}                               & \multicolumn{1}{c|}{\cellcolor[HTML]{FFFFFF}1.18}               & \multicolumn{1}{c|}{\cellcolor[HTML]{FFFFFF}1.03}               & 1.00               & \multicolumn{1}{c|}{\cellcolor[HTML]{FFFFFF}25.31}                              & \multicolumn{1}{c|}{\cellcolor[HTML]{FFFFFF}27.16}              & \multicolumn{1}{c|}{\cellcolor[HTML]{FFFFFF}36.42}              & 38.27              & \multicolumn{1}{c|}{\cellcolor[HTML]{FFFFFF}2257.50}                            & 94.44                                                                                                    \\ \hline
\rowcolor[HTML]{FFFFFF} 
\cellcolor[HTML]{EDF9FA}\textbf{multiplier}                   & \multicolumn{1}{c|}{\cellcolor[HTML]{FFFFFF}2.44}                              & \multicolumn{1}{c|}{\cellcolor[HTML]{FFFFFF}1.23}                               & \multicolumn{1}{c|}{\cellcolor[HTML]{FFFFFF}1.22}               & \multicolumn{1}{c|}{\cellcolor[HTML]{FFFFFF}1.14}               & 1.12               & \multicolumn{1}{c|}{\cellcolor[HTML]{FFFFFF}49.59}                              & \multicolumn{1}{c|}{\cellcolor[HTML]{FFFFFF}50.00}              & \multicolumn{1}{c|}{\cellcolor[HTML]{FFFFFF}53.28}              & 54.10              & \multicolumn{1}{c|}{\cellcolor[HTML]{FFFFFF}498.15}                             & 127.03                                                                                                   \\ \hline
\rowcolor[HTML]{FFFFFF} 
\cellcolor[HTML]{EDF9FA}\textbf{log2}                         & \multicolumn{1}{c|}{\cellcolor[HTML]{FFFFFF}2.25}                              & \multicolumn{1}{c|}{\cellcolor[HTML]{FFFFFF}1.16}                               & \multicolumn{1}{c|}{\cellcolor[HTML]{FFFFFF}1.20}               & \multicolumn{1}{c|}{\cellcolor[HTML]{FFFFFF}1.11}               & 1.06               & \multicolumn{1}{c|}{\cellcolor[HTML]{FFFFFF}48.44}                              & \multicolumn{1}{c|}{\cellcolor[HTML]{FFFFFF}46.67}              & \multicolumn{1}{c|}{\cellcolor[HTML]{FFFFFF}50.67}              & 52.89              & \multicolumn{1}{c|}{\cellcolor[HTML]{FFFFFF}428.31}                             & 148.38                                                                                                   \\ \hline
\rowcolor[HTML]{F2FAED} 
\textbf{Average}                                              & \multicolumn{1}{c|}{\cellcolor[HTML]{F2FAED}\textbf{2.43}}                     & \multicolumn{1}{c|}{\cellcolor[HTML]{F2FAED}\textbf{1.27}}                      & \multicolumn{1}{c|}{\cellcolor[HTML]{F2FAED}\textbf{1.26}}      & \multicolumn{1}{c|}{\cellcolor[HTML]{F2FAED}\textbf{1.15}}      & \textbf{1.07}      & \multicolumn{1}{c|}{\cellcolor[HTML]{F2FAED}\textbf{45.46}}                     & \multicolumn{1}{c|}{\cellcolor[HTML]{F2FAED}\textbf{45.72}}     & \multicolumn{1}{c|}{\cellcolor[HTML]{F2FAED}\textbf{50.88}}     & \textbf{54.09}     & \multicolumn{1}{c|}{\cellcolor[HTML]{F2FAED}\textbf{420.79}}                    & \textbf{66.25}                                                                                          \\ \hline
\bottomrule
\end{tabular}
}
\raggedright \footnotesize{Note: \textbf{`X\% Mask'} denotes X\% of total number of leaky gates. The computational time for \polarai~with varying mask sizes is similar.}
\end{threeparttable}
\vspace{-0.2in}
\end{table*}
\begin{table}[!htbp]
\centering
\caption{Comparison among different ML models used in \polarai. Values indicate leakage reduction in \%.}
\label{table:XDFS_AI_Models}


\begin{tabular}{|
>{\columncolor[HTML]{EDF9FA}}c |c|c|c|}
\hline
\cellcolor[HTML]{ECF4FF}\textbf{Designs} & \cellcolor[HTML]{ECF4FF}\textbf{Random Forest} & \cellcolor[HTML]{ECF4FF}\textbf{XGBoost} & \cellcolor[HTML]{ECF4FF}\textbf{AdaBoost} \\ \hline
\textbf{des3}                               & 33.73                                     & 41.35                                     & 61.65                                      \\ \hline
\textbf{arbiter}                            & 29.36                                     & 43.02                                     & 38.03                                      \\ \hline
\textbf{sin}                                & 32.48                                     & 56.89                                     & 57.87                                      \\ \hline
\textbf{md5}                                & 43.97                                     & 54.70                                     & 58.70                                      \\ \hline
\textbf{voter}                              & 58.67                                     & 62.39                                     & 65.88                                      \\ \hline
\textbf{square}                             & 22.37                                     & 42.96                                     & 58.84                                      \\ \hline
\textbf{sqrt}                               & 39.02                                     & 53.57                                     & 60.53                                      \\ \hline
\textbf{div}                                & 68.69                                     & 74.36                                     & 48.25                                      \\ \hline
\textbf{memctrl}                            & 18.95                                     & 28.16                                     & 38.27                                      \\ \hline
\textbf{multiplier}                         & 57.09                                     & 55.09                                     & 54.10                                      \\ \hline
\textbf{log2}                               & 57.37                                     & 53.88                                     & 52.89                                      \\ \hline
\cellcolor[HTML]{F2FAED}\textbf{Average}    & \cellcolor[HTML]{F2FAED}\textbf{41.97}    & \cellcolor[HTML]{F2FAED}\textbf{51.49}    & \cellcolor[HTML]{F2FAED}\textbf{54.09}     \\ \hline \hline
\end{tabular}
\\ \raggedright \footnotesize{Note: We observe nominal differences in computational time across models.}
\end{table}

\section{Results and XAI Analysis}\label{sec:results} 
In this section, we analyze the effectiveness of the proposed framework \polarai~in identifying and mitigating leakage from open-source designs with varying complexities and leveraging the SHAP framework to mitigate leakage through masking against power side-channel attacks.

\begin{table*}[!htbp]
\centering
\captionsetup{justification=centering}
\caption{Comparison of area, power, and delay overheads between VALIANT \cite{valiant} and \polarai\textsuperscript{\S}.}
\label{table:overhead}
\begin{threeparttable}
\resizebox{\textwidth}{!}{%
\begin{tabular}{|c|ccc|ccc|ccc|ccc|}
\hline
\rowcolor[HTML]{ECF4FF} 
\cellcolor[HTML]{ECF4FF}                                   & \multicolumn{3}{c|}{\cellcolor[HTML]{ECF4FF}\textbf{Original}}                                                                                               & \multicolumn{3}{c|}{\cellcolor[HTML]{ECF4FF}\textbf{VALIANT (x Original\textsuperscript{$\diamondsuit$})}}                                                                      & \multicolumn{3}{c|}{\cellcolor[HTML]{ECF4FF}\textbf{\polarai\textsuperscript{\S} (x Original\textsuperscript{$\diamondsuit$})}}                                                                  & \multicolumn{3}{c|}{\cellcolor[HTML]{ECF4FF}\textbf{Reduction (\%) in \polarai}}                                                       \\ \cline{2-13} 
\rowcolor[HTML]{EFEFEF} 
\multirow{-2}{*}{\cellcolor[HTML]{ECF4FF}\textbf{Designs}} & \multicolumn{1}{c|}{\cellcolor[HTML]{EFEFEF}\textbf{Area (µM²)}} & \multicolumn{1}{c|}{\cellcolor[HTML]{EFEFEF}\textbf{Power (mW)}} & \textbf{Delay (ns)}   & \multicolumn{1}{c|}{\cellcolor[HTML]{EFEFEF}\textbf{Area}} & \multicolumn{1}{c|}{\cellcolor[HTML]{EFEFEF}\textbf{Power}} & \textbf{Delay}        & \multicolumn{1}{c|}{\cellcolor[HTML]{EFEFEF}\textbf{Area}} & \multicolumn{1}{c|}{\cellcolor[HTML]{EFEFEF}\textbf{Power}} & \textbf{Delay}        & \multicolumn{1}{c|}{\cellcolor[HTML]{EFEFEF}\textbf{Area}}  & \multicolumn{1}{c|}{\cellcolor[HTML]{EFEFEF}\textbf{Power}} & \textbf{Delay}        \\ \hline
\cellcolor[HTML]{EDF9FA}\textbf{des3}                      & \multicolumn{1}{c|}{9083.31}                                            & \multicolumn{1}{c|}{2.73}                                             & \multicolumn{1}{c|}{1.3} & \multicolumn{1}{c|}{3.7}                                      & \multicolumn{1}{c|}{3.1}                                       & \multicolumn{1}{c|}{1.6} & \multicolumn{1}{c|}{2.4}                                      & \multicolumn{1}{c|}{1.8}                                       & \multicolumn{1}{c|}{1.2} & \multicolumn{1}{c|}{35.14}                                       & \multicolumn{1}{c|}{41.94}                                       & \multicolumn{1}{c|}{25.00} \\ \hline
\cellcolor[HTML]{EDF9FA}\textbf{arbiter}                   & \multicolumn{1}{c|}{10310.05}                                    & \multicolumn{1}{c|}{1.61}                                         & 1.23                  & \multicolumn{1}{c|}{3.1}                                  & \multicolumn{1}{c|}{3.3}                                   & 3.3                  & \multicolumn{1}{c|}{2.1}                                  & \multicolumn{1}{c|}{2}                                   & 2.1                  & \multicolumn{1}{c|}{32.26}                                  & \multicolumn{1}{c|}{39.39}                                  & 36.36                 \\ \hline
\cellcolor[HTML]{EDF9FA}\textbf{sin}                       & \multicolumn{1}{c|}{13421.04}                                    & \multicolumn{1}{c|}{9.12}                                         & 7.98                  & \multicolumn{1}{c|}{3.6}                                  & \multicolumn{1}{c|}{3.1}                                   & 3.1                  & \multicolumn{1}{c|}{2.3}                                  & \multicolumn{1}{c|}{1.6}                                   & 1.9                  & \multicolumn{1}{c|}{36.11}                                  & \multicolumn{1}{c|}{48.39}                                  & 38.71                 \\ \hline
\cellcolor[HTML]{EDF9FA}\textbf{md5}                       & \multicolumn{1}{c|}{24217.29}                                            & \multicolumn{1}{c|}{0.94}                                             & \multicolumn{1}{c|}{0.12} & \multicolumn{1}{c|}{5.1}                                      & \multicolumn{1}{c|}{3.8}                                       & \multicolumn{1}{c|}{1.4} & \multicolumn{1}{c|}{3.1}                                      & \multicolumn{1}{c|}{1.7}                                       & \multicolumn{1}{c|}{1.1} & \multicolumn{1}{c|}{39.22}                                       & \multicolumn{1}{c|}{55.26}                                       & \multicolumn{1}{c|}{21.43} \\ \hline
\cellcolor[HTML]{EDF9FA}\textbf{voter}                     & \multicolumn{1}{c|}{28090.42}                                    & \multicolumn{1}{c|}{17.25}                                        & 2.53                  & \multicolumn{1}{c|}{4.2}                                  & \multicolumn{1}{c|}{3.8}                                   & 4.5                  & \multicolumn{1}{c|}{2.5}                                  & \multicolumn{1}{c|}{2.4}                                   & 2.9                  & \multicolumn{1}{c|}{40.48}                                  & \multicolumn{1}{c|}{36.84}                                  & 35.56                 \\ \hline
\cellcolor[HTML]{EDF9FA}\textbf{square}                    & \multicolumn{1}{c|}{51117.09}                                    & \multicolumn{1}{c|}{32.98}                                        & 9.84                  & \multicolumn{1}{c|}{4.7}                                  & \multicolumn{1}{c|}{3.8}                                   & 3.2                  & \multicolumn{1}{c|}{2.7}                                  & \multicolumn{1}{c|}{2.3}                                   & 1.2                  & \multicolumn{1}{c|}{42.55}                                  & \multicolumn{1}{c|}{39.47}                                  & 62.50                 \\ \hline
\cellcolor[HTML]{EDF9FA}\textbf{sqrt}                      & \multicolumn{1}{c|}{42921.71}                                    & \multicolumn{1}{c|}{83.28}                                        & 279.99                & \multicolumn{1}{c|}{5.3}                                  & \multicolumn{1}{c|}{3.8}                                   & 3                  & \multicolumn{1}{c|}{3.2}                                  & \multicolumn{1}{c|}{2.3}                                   & 1.9                  & \multicolumn{1}{c|}{39.62}                                  & \multicolumn{1}{c|}{39.47}                                  & 36.67                 \\ \hline
\cellcolor[HTML]{EDF9FA}\textbf{div}                       & \multicolumn{1}{c|}{44048.97}                                    & \multicolumn{1}{c|}{12.59}                                        & 196.90                & \multicolumn{1}{c|}{3.1}                                  & \multicolumn{1}{c|}{2.9}                                   & 2.2                  & \multicolumn{1}{c|}{2.5}                                  & \multicolumn{1}{c|}{2.2}                                   & 1.7                  & \multicolumn{1}{c|}{19.35}                                  & \multicolumn{1}{c|}{24.14}                                  & 22.73                 \\ \hline
\cellcolor[HTML]{EDF9FA}\textbf{memctrl}                   & \multicolumn{1}{c|}{42921.71}                                    & \multicolumn{1}{c|}{8.21}                                         & 3.55                  & \multicolumn{1}{c|}{4}                                  & \multicolumn{1}{c|}{4.2}                                   & 2.6                  & \multicolumn{1}{c|}{2.2}                                  & \multicolumn{1}{c|}{2.1}                                   & 1.9                  & \multicolumn{1}{c|}{45.00}                                  & \multicolumn{1}{c|}{50.00}                                  & 26.92                 \\ \hline
\cellcolor[HTML]{EDF9FA}\textbf{multiplier}                & \multicolumn{1}{c|}{66185.85}                                    & \multicolumn{1}{c|}{56.62}                                        & 10.13                 & \multicolumn{1}{c|}{2.4}                                  & \multicolumn{1}{c|}{1.9}                                   & 2                  & \multicolumn{1}{c|}{2.1}                                  & \multicolumn{1}{c|}{1.5}                                   & 1.7                  & \multicolumn{1}{c|}{12.50}                                   & \multicolumn{1}{c|}{21.05}                                   & 21.68                \\ \hline
\cellcolor[HTML]{EDF9FA}\textbf{log2}                      & \multicolumn{1}{c|}{77395.08}                                    & \multicolumn{1}{c|}{66.95}                                        & 16.55                 & \multicolumn{1}{c|}{3.9}                                  & \multicolumn{1}{c|}{3.2}                                   & 3.4                  & \multicolumn{1}{c|}{2.4}                                  & \multicolumn{1}{c|}{1.6}                                   & 2.1                  & \multicolumn{1}{c|}{38.46}                                   & \multicolumn{1}{c|}{50.00}                                   & 38.24                \\ \hline
\rowcolor[HTML]{F2FAED} 
\textbf{Average}                                           & \multicolumn{1}{c|}{\cellcolor[HTML]{F2FAED}\textbf{37246.59}}   & \multicolumn{1}{c|}{\cellcolor[HTML]{F2FAED}\textbf{26.57}}       & \textbf{48.19}        & \multicolumn{1}{c|}{\cellcolor[HTML]{F2FAED}\textbf{3.92}} & \multicolumn{1}{c|}{\cellcolor[HTML]{F2FAED}\textbf{3.35}}  & \textbf{2.75}         & \multicolumn{1}{c|}{\cellcolor[HTML]{F2FAED}\textbf{2.50}} & \multicolumn{1}{c|}{\cellcolor[HTML]{F2FAED}\textbf{1.95}}  & \textbf{1.79}         & \multicolumn{1}{c|}{\cellcolor[HTML]{F2FAED}\textbf{34.61}} & \multicolumn{1}{c|}{\cellcolor[HTML]{F2FAED}\textbf{40.54}} & \textbf{33.25}        \\ \hline
\bottomrule
\end{tabular}
}
\raggedright \footnotesize{\textbf{\S} We utilize \polarai~w/ 50\% Mask for comparison with VALIANT, achieving a comparable leakage reduction while masking half the number of gates.\\}
\raggedright \footnotesize{\textbf{$\diamondsuit$} Overheads are reported as $\textsf{x}$ times original value.}
\end{threeparttable}
\vspace{-0.1in}
\end{table*}\
\begin{table*}[!htbp]
\caption{Power side-channel mitigation rules generated via the \polarai~framework (AdaBoost Model).}
\centering
\label{table:extractedRules_sc}
\small\addtolength{\tabcolsep}{-1pt}
\begin{tabular}{|
>{\columncolor[HTML]{EDF9FA}}c |
>{\columncolor[HTML]{FFFFFF}}c |
>{\columncolor[HTML]{FFFFFF}}c |}
\hline
\cellcolor[HTML]{ECF4FF}\textbf{Rules} &
  \cellcolor[HTML]{ECF4FF}\textbf{As long as} &
  \cellcolor[HTML]{ECF4FF}\textbf{Procedure} \\ \hline
\cellcolor[HTML]{EDF9FA}\textbf{A} &
  \cellcolor[HTML]{FFFFFF}\begin{tabular}[c]{@{}c@{}}G4 = NAND \&\& G5 = AND \&\&  G4 (NAND) and G5 (AND) are not connected \&\&\\ G6 = NAND \&\& G10 = NAND \&\& G8 (NOT) and G9 (NAND) are connected\end{tabular} & 
  \begin{tabular}[c]{@{}c@{}}
  Select \& Replace \\ with masking gate \end{tabular}\\ \hline
\textbf{B} &
  \cellcolor[HTML]{FFFFFF}\begin{tabular}[c]{@{}c@{}}G5 = AND \&\&  G4 = NOT \&\& G7 = OR \&\& G4 (NOT) and G5 (AND) are connected blue{}\\ \&\&  G7 (OR) and G8 (NOT) are connected\end{tabular} &
  Do not Mask \\ \hline

\end{tabular}

\end{table*}

\subsection{Experiment Configuration}
We utilize knowledge obtained from training on smaller designs to generalize to larger and entirely unseen designs, leveraging the transfer learning approach. The model is trained on smaller open-source designs that can adapt to larger designs, capturing patterns that can be transferred for practical application of transfer learning. We used six designs from the ISCAS-85 benchmark suite \cite{iscas85} synthesized using Synopsys Design Compiler (DC) for training and simulated using 10,000 traces for leakage measurement by TVLA analysis. Selecting smaller designs helps reduce the model's training time ($\sim$40 minutes). The key parameters are: $\mathcal{M}_{size} = 200$, $\mathcal{L} = 7$ (considering 7 neighboring gates), $itr = 100$, and $\theta_r = 0.70$ (indicating a leakage reduction of $70\%$ or more as good masking). The rationale behind selecting the value of $\theta_r$ to 0.70 is that selecting higher values lead to significant data imbalance, which could cause the model to underfit and hinder its ability to generalize effectively. The evaluation phase includes larger and entirely different designs with varying $\mathcal{M}_{size}$ from EPFL \cite{epfl} and MIT-CEP \cite{mit_cep} benchmark suite that are distinct from training designs. 


\subsection{Comparing ML Models for Leakage Reduction}


In this work, we explore several machine-learning techniques to develop improved masked designs that mitigate side-channel vulnerabilities. Table~\ref{table:XDFS_AI_Models} compares the leakage reduction by Random Forest, XGBoost, and AdaBoost models for $\mathcal{L} = 7$, $\theta_r = 0.7$, and $\mathcal{M}_{size}$ denoting number of leaky gates identified by TVLA. We applied SMOTE \cite{chawla2002smote} for Random Forest and employed weighted training for XGBoost and AdaBoost models for handling imbalance that occurred due to $\theta_r$. The learning rate ($\alpha$) for both XGBoost and AdaBoost was set to 0.01. AdaBoost outperforms the other models with a 54.09\% leakage reduction on average, making it the chosen model for the remaining experiments.




\begin{figure}[!htbp]
    \centering
    \subfloat[]{\includegraphics[width=0.9\columnwidth]{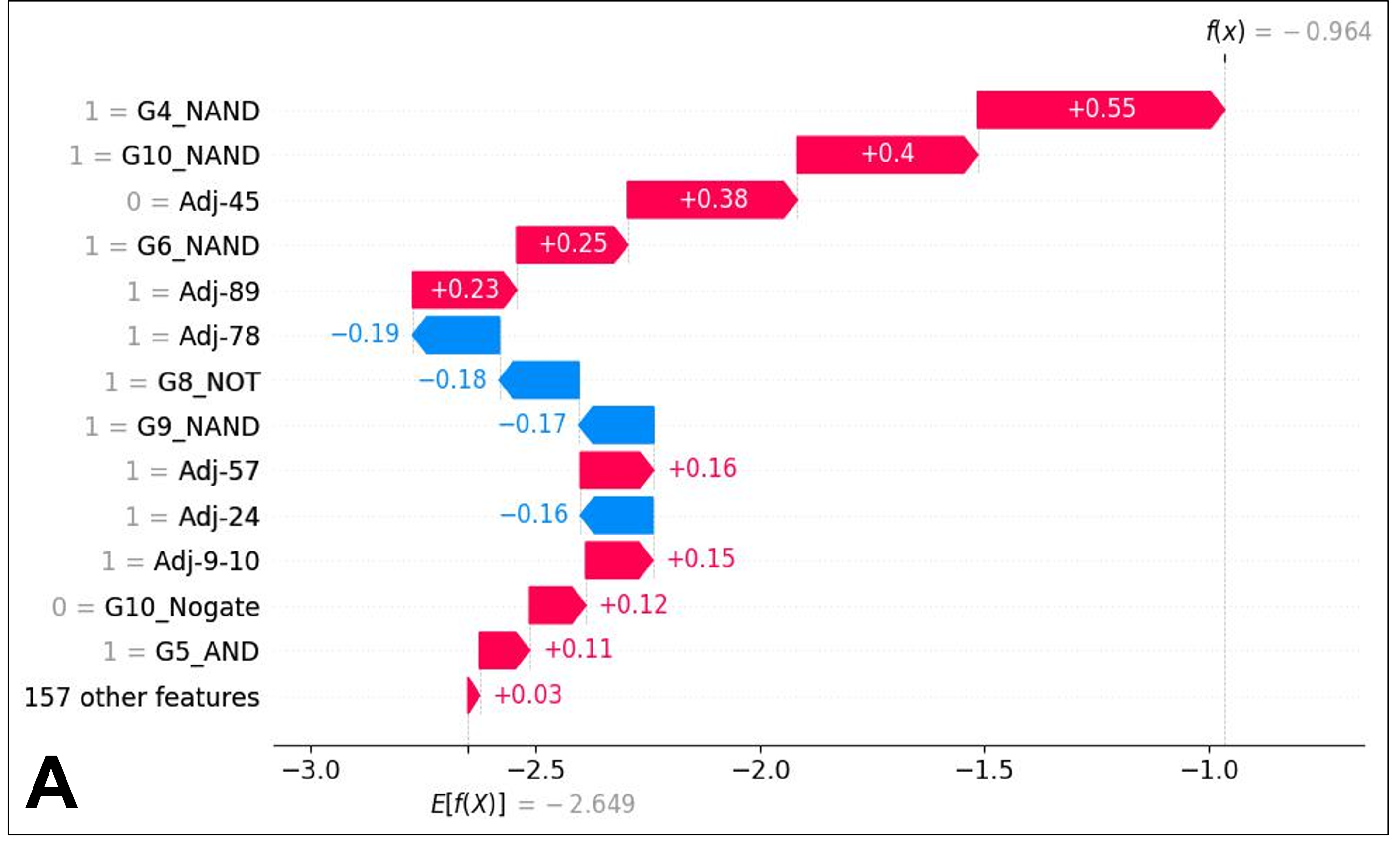}}
    \vfill
    \subfloat[]{\includegraphics[width=0.9\columnwidth]{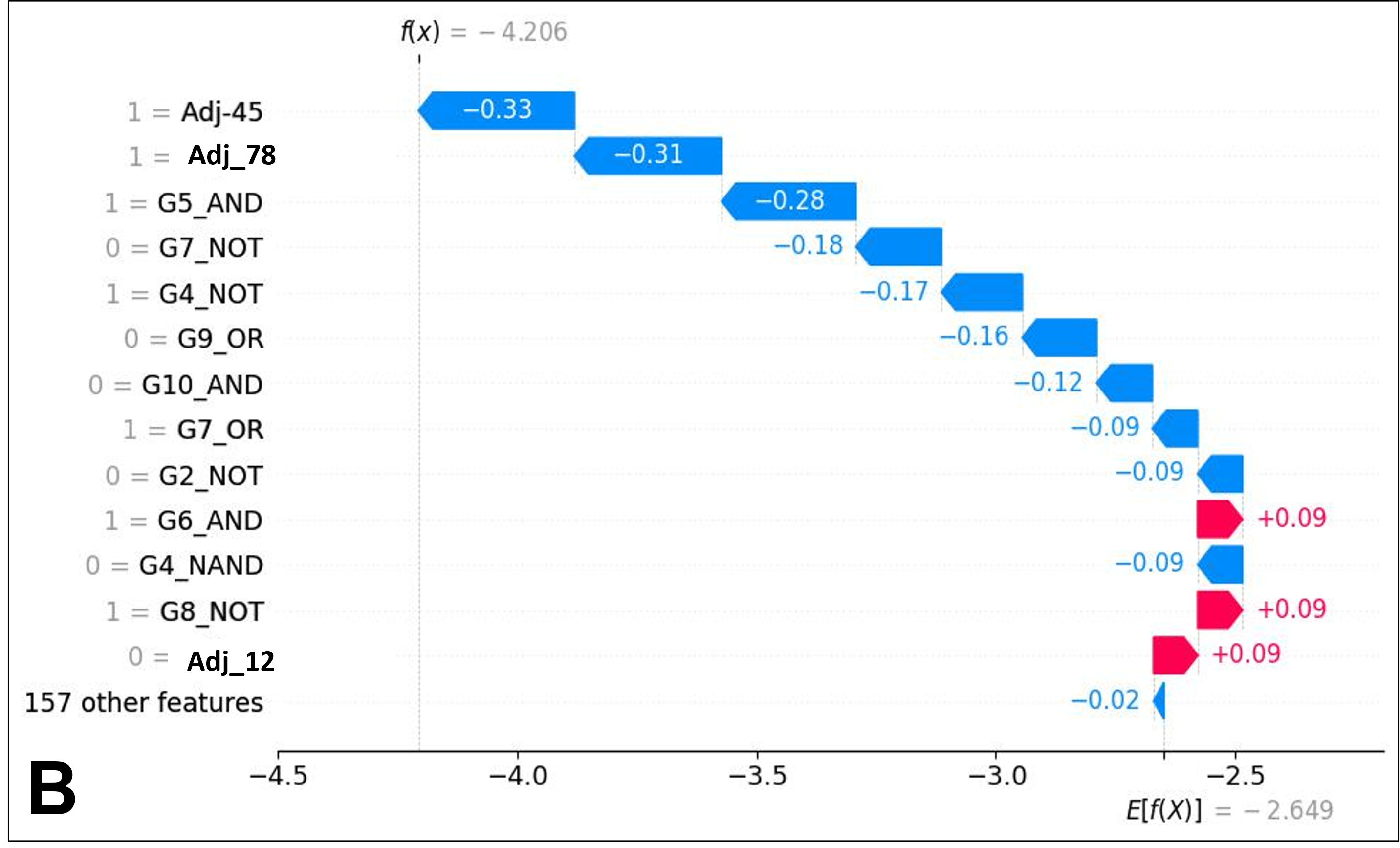}}
    \caption{SHAP waterfall plots generated by \polarai~Adaboost (ADB) model for power side channel defense.}%
    \label{fig:add_rules_sc}
\end{figure}
\begin{figure}[!htb]
\centering

\includegraphics[width=\columnwidth]{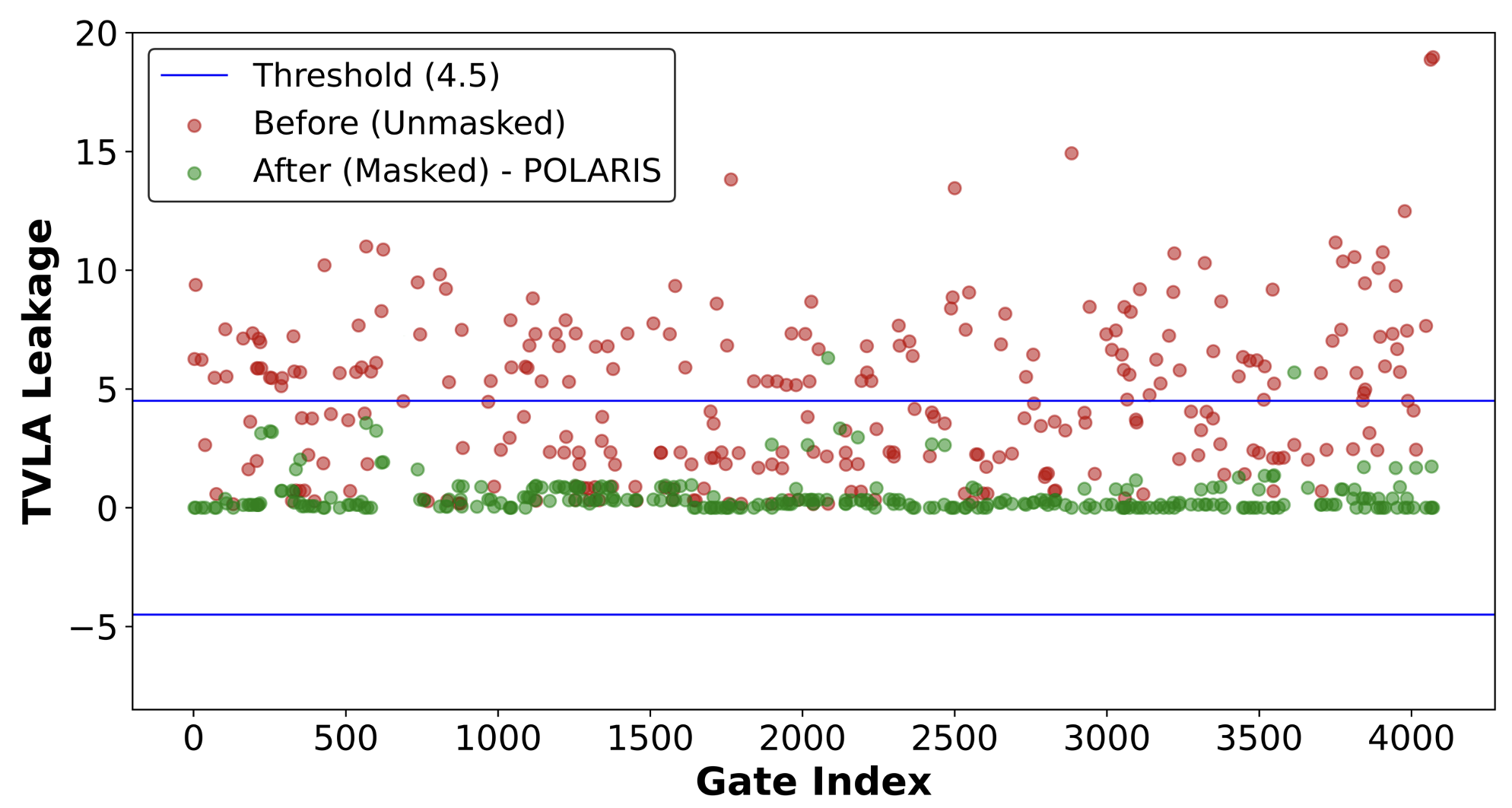}
\caption{TVLA values before and after masking in \textit{des3} design. Gates exceeding threshold ($\pm$4.5) are considered as \textit{leaky}.\label{fig:threshold}}

\end{figure}

\vspace{-0.2in}
\subsection{Comparison with State-of-the-Art}

Table~\ref{table:core_result} compares the performance of VALIANT framework \cite{valiant} and \polarai~in terms of leakage per gate, total leakage reduction, and execution time. \polarai~is tested with 50\%, 75\%, and full (100\%) masking sizes. Even with only 50\% masking, \polarai~achieves a 45.72\% leakage reduction, surpassing VALIANT’s 45.46\% with full masking. At 75\% and full masking, \polarai~reduces leakage by 50.88\% and 54.09\%, respectively, outperforming VALIANT. Additionally, \polarai~operates $6\mathsf{x}$ faster than VALIANT, demonstrating its efficiency in leakage mitigation. Table~\ref{table:overhead} compares area ($\mu M^2$), power ($mW$), and delay ($ns$) overheads of the unmasked design (Original) with the respective masked designs produced by VALIANT and \polarai. As we can observe, \polarai~reduces area by 34.61\%, power by 40.54\%, and delay by 33.25\% compared to VALIANT while also achieving greater leakage reduction. These results highlight the effectiveness of \polarai~in mitigating side-channel vulnerabilities with reduced resource overhead.

\subsection{Explainability: \polarai~Extracted Rules}

Using SHAP algorithms, we interpret the decision-making process for individual samples by examining waterfall plots in Fig.~\ref{fig:add_rules_sc}, which illustrate the contribution of each feature to the prediction of the target variable. Here, $x$ represents a selected observation, $f(x)$ is the model's predicted value for this input, and $E[f(x)]$ denotes the expected value of the target variable, essentially the average prediction across all observations. The length of each bar in the plot indicates the SHAP value (Fig.~\ref{fig:add_rules_sc}). \polarai~can extract rules for leakage reduction; these rules are summarized in Table~\ref{table:extractedRules_sc}.


\subsection{Additional Analysis \& Discussion}

To visualize the effectiveness of \polarai~(see Fig.~\ref{fig:threshold}), we performed TVLA analysis for all gates in the \textit{des3} design \textit{before} (\tikzcircle[fill=purple!80]{3pt}) and \textit{after \polarai~masking} (\tikzcircle[fill=OliveGreen]{3pt}). The performance of \polarai~can be improved by including a wider variety of designs in the training process, enabling \polarai~to better comprehend design diversity. 
\polarai~can also be extended to support other masking gates (e.g., DOM \cite{dom}), providing a flexible and robust solution for addressing power side-channel leakage mitigation with improved performance. 


\vspace{-0.1in} 
\section{Conclusion}
\label{sec:conclusion}
Power side-channel leakage of sensitive on-chip data is a major security threat. Existing mitigation frameworks face scalability issues (limited data, TVLA reliance, slow runtime) and lack effectiveness. This paper presents the \polarai, a DFS framework that can be integrated into commercial ASIC design flow to generate design-specific rules using XAI for low-overhead power side-channel mitigation. \polarai~is fast and utilizes a synthetic data generation scheme to bypass data-related concerns associated with most AI frameworks. We observe significant performance gain over state-of-the-art solutions in terms of speed, leakage reduction, and design overhead. Future works will make the \polarai~framework more efficient and extend it to other side-channels (EM/timing). 

\vspace{-0.12in}
\section{Acknowledgment}
This material is supported by the National Science Foundation (NSF) under Grant No. 2350363 and Grant No. 2316399.

\bibliographystyle{IEEEtran}
\bibliography{IEEEabrv,XDFS}

\end{document}